\documentclass[aps,prb,preprint,groupedaddress]{revtex4}
\usepackage{graphicx}

\begin{document}


\title{Tunable Band Gap of Boron Nitride Interfaces under Uniaxial Pressure}


\author{Elizane E. Moraes, Ta\'{i}se M. Manhabosco, Alan B. de Oliveira, and Ronaldo J. C. Batista}
\affiliation{Departamento de F\'{\i}sica,
Universidade Federal de Ouro Preto, Ouro Preto, MG, 35400-000, Brazil}


\date{\today}

\begin{abstract}
In this work we show, by means of a density functional theory formalism, that the simple physical contact between
hydrogen terminated boron nitride surfaces gives rise to a metallic interface with free carries of opposite sign
at each surface. A band gap can be induced by applying uniaxial pressure. The size of the band gap changes continuously
from zero up to 4.4~eV with increasing pressure, which is understood in terms of the interaction between surface states. 
Due to the high thermal conductivity of cubic boron nitride and the coupling
between band gap and applied pressure, such tunable band gap interfaces may be used in high stable electronic and electromechanical devices.
In addition, the spacial separation of charge carries at the interface may lead to photovoltaic applications. 

\end{abstract}


\maketitle

Diamond is commonly used as an abrasive material in technological and industrial applications due to its hardness.
Other intrinsic properties of diamond have increased its range of applications from biotechnology to electronics
\cite{MM2009,2MM2009, DK2011, MFMRM2011}. In electronics, in particular, its large electronic band gap and exceptionally
high thermal conductivity led to the production of highly stable solid state transistors (FETs) based on hydrogen
terminated diamond \cite{DK2011,MFMRM2011,TUATK2001}. Due to the similarity between the isoelectronic C-C and B-N bonds, boron and nitrogen
form a material analogous to the carbon diamond, namely, cubic boron nitride (cBN). cBN presents hardness, band gap and
thermal conductivity very similar to those of carbon diamond. In addition, it is also possible to produce hydrogen
terminated cBN \cite{rofrh2000} surfaces. Nevertheless, the C-H is not isoeletronic to B-H and N-H bonds, thus
hydrogen terminated diamond and hydrogen terminated cBN must present distinct electronic and electrochemical properties.
For example, the band gap of thin hydrogen terminated cBN
films depends on the number of BN layers due to the differences between B-H and N-H terminated surfaces \cite{zg2012}. 
On the other hand, the band gap of thin diamond films is constant.

In this work, we apply a density functional formalism to investigate interfaces formed by the physical contact between hydrogen
terminated boron nitride surfaces. We found that those interfaces present a band gap that can be continuously controlled by
applied uniaxial pressure, which is explained in terms of the interaction between surface states due to the B-H and N-H terminations. 
The  gap of the system can be modified from zero up to roughly 4.4~eV (a value of band gap close to that calculated for 
cBN within density functional theory approach). At null values of pressure the interface presents a metallic behavior 
with electrons at N-H terminated surface and
holes a B-H terminated surface. The semiconductor, metallic and insulating behavior of such interfaces may allow their
use in electronics as high stable devices. The coupling between applied pressure and the band gap, on the other hand,
may permit the use of such interfaces in electromechanical devices. 
Besides, since the free carries are spatially separated in those interfaces and the cBN is transparent for
visible light, they could be also used in photovoltaic applications.

The applied first-principles methodology is based on the Density Functional Theory (DFT) as implemented
in the SIESTA program \cite{siesta}. We used the Generalized Gradient
Approximation (GGA) as parametrized in the Perdew-Burke-Ernzerhof scheme (PBE) \cite{pbe} for the exchange-correlation functional.
 The ionic core potentials were represented
by norm-conserving scalar relativistic Troullier-Martins\cite{martins} pseudopotentials
in Kleinman-Bylander nonlocal form \cite{kleinman}.
The fineness of the real-space grid integration was defined by a minimal energy
cutoff of 150~Ry \cite{josemesh}. The geometries were fully optimized using the conjugate gradient
algorithm \cite{payne} until all the force components were smaller than 0.02 eV/\AA.
The Kohn-Sham (KS) eigenfunctions were expanded as linear
combination of pseudo atomic orbitals of finite range consisting of double-zeta
radial functions per angular momentum plus polarization orbitals (DZP).

\begin{figure}
 \centering
\includegraphics[clip=true,scale=0.35]{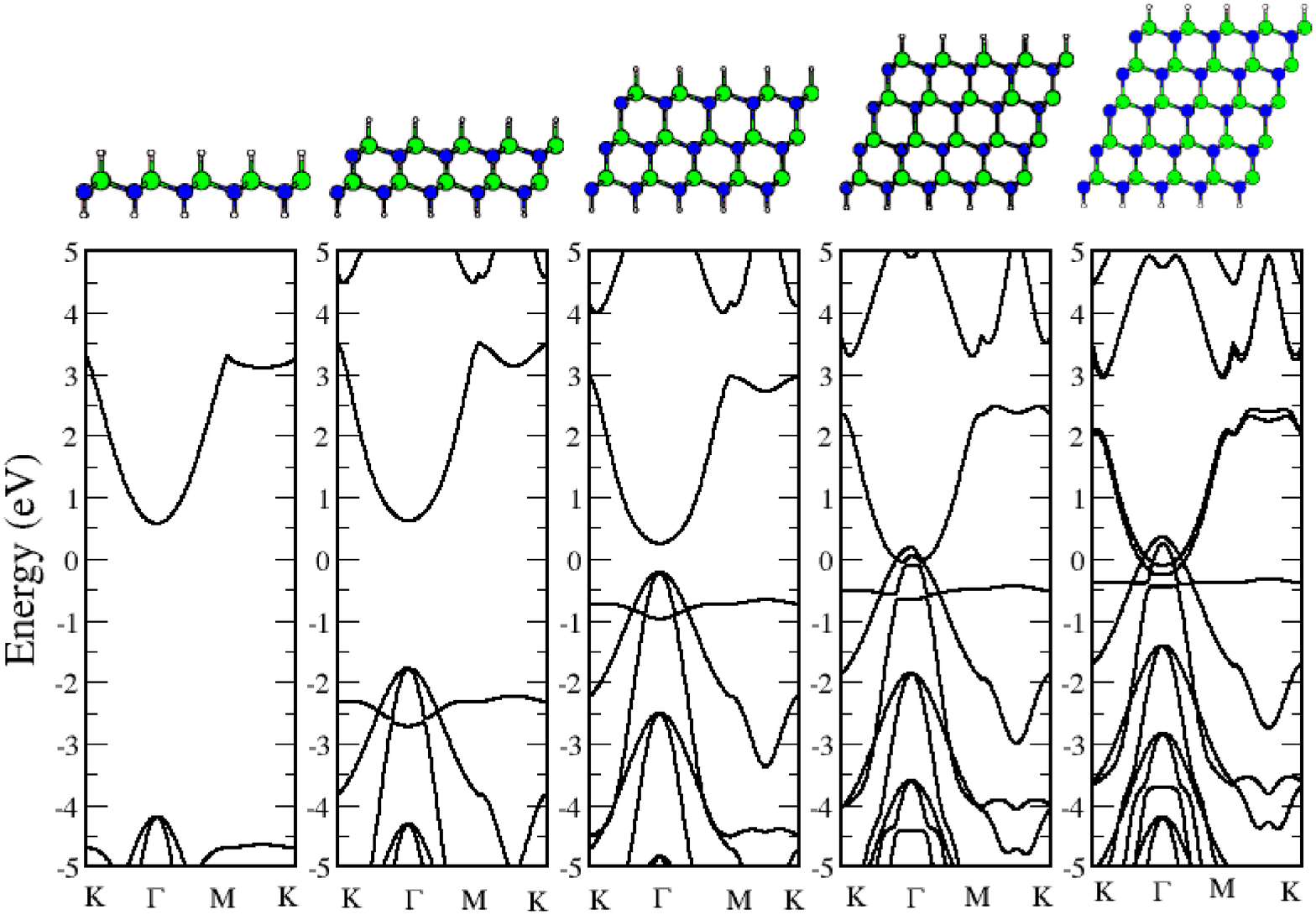}
 \caption{Bands structure of thin boron nitride nanofilms. From left to right the number of layers increases from one up to five.
As the number of layers decreases the overlap between orbitals of atoms at each surface increases giving rise to band gaps.}
 \label{figure1}
\end{figure}

As a first step to understand the electronic properties of cBN interfaces we investigated few-layered hydrogenated
(111) BN films. The top panels of Fig. 1 show the geometries of the few-layered films whose number of layers changes
from one up to five while the bottom panels show the respective electronic structure. It is clear that
there is  a dependence of the band gap with the number of layers, which can be understood in terms of differences between B-H and N-H $sp^{3}$ terminations.

The C-C $sp^{3}$ bonds in diamond share a pair of
electrons (satisfying then the octet rule) which leads to a wide band gap between valence and conducting states.
Since boron and nitrogen are the carbon left and right neighbors in the periodic table of elements, respectively, the B-N bond is isoelectronic to C-
C bond. Therefore, the B-N $sp^{3}$ bonds in cBN also share a pair of electrons where the excess of electrons
of nitrogen orbitals are precisely offset by the lack of electrons in boron orbitals. As a result, cBN presents a band gap as wide as 
that of carbon diamond. 
In the N-H and B-H bonds on hydrogen terminated cBN, on the other
hand, the H atoms do not compensate the excess or lack of electrons on N and B orbitals in cBN surfaces, which results in surface
states within the large band gap of cBN. In a sufficiently thick film the interaction between those surfaces is
negligible. This is the case of films with four layers or more, in which there is no overlap
between N-H and B-H orbitals (indeed, the distance between N and B atoms at both ends in the 4-layer BN film is 7.12~\AA, 
that is greater than two times the cutoff radius of N and B  orbitals, 2.5~\AA). In those cases, the edge of the occupied surface state localized
at the B-H end is above the edge of the unoccupied surface state localized at the N-H end (the fourth and fifth panels of Fig. 1 show 
a superposition of non-interacting surface states). Thus, the N-H terminated surface can transfer charge to the B-H
terminated surface if a physical contact is provided. 
In a very recent work \cite{nos}, we have investigated by means of theory and experiment a similar case in which
the physical contact between semiconducting carbon nanotubes (SCNT) and diamond surfaces -- which are large gap insulators --
allows a metallic response. In that case, the top of valence band of diamond surface was above the bottom of the conduction
band of SCNT.
 
A physical contact between N-H and B-H terminated surface occurs in films with less than 3 layers since there is an overlap
between orbitals of B and N atoms at different surfaces. In a single layer film (left panel of Fig. 1) the overlap between N 
and B orbitals is maximum, then the excess or lack of electrons at N-H and B-H surfaces is canceled out. 
As a result the one layer cBN film presents a band gap very similar to that
of cBN bulk. The overlap between orbitals of the atoms at different surfaces decreases with film thickness which leads to
values of band gaps shown in the second and third panels of Fig. 1.

\begin{figure}
 \centering
\includegraphics[clip=true,scale=0.35]{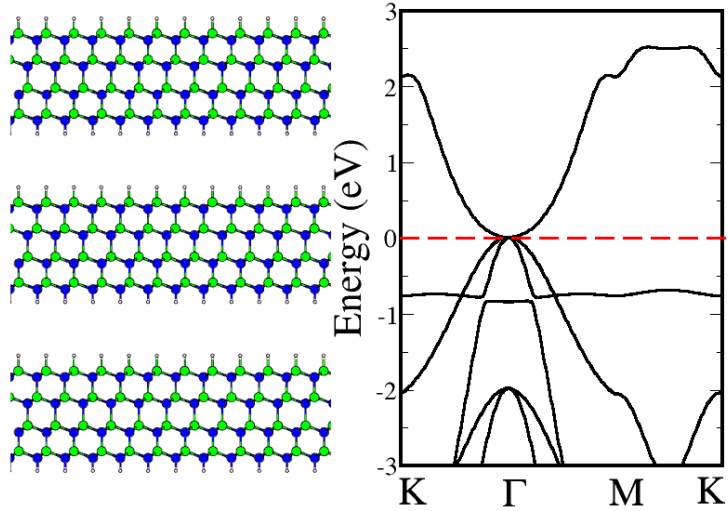}
 \caption{Left panel: interfaces formed by the physical contact between N-H and B-H terminated cBN four-layered films, only the
first neighbors periodical images are shown. Right panel: bands structure of the interfaces shown in the left panel.}
 \label{figure2}
\end{figure}

The results discussed above strongly suggest that the overlap between orbitals of the atoms of B-H and N-H cBN surface
determine the band gap since it affects charge distribution that gives rise to the surface states.
The overlap between the orbitals of atoms at different surfaces can be continuously controlled by the physical contact 
between surfaces as shown in the left panel of Fig. 2. We have simulated the physical contact between B-H and N-H terminated cBN surfaces 
by performing calculations on 4-layer cBN nanofilms in which he distance between successive periodical 
images allows the interaction between B-H and N-H terminated surfaces.
The right panel of Fig. 2 shows the band structure of an interface formed by the simple deposition of a H terminated cBN surface onto its 
counterpart, at null values of uniaxial pressure. It is possible to see that the band structure of such an interface differs from the 
band structure of an isolated four layer film (fourth panel of Fig. 1). Such a difference is due to a charge transfer between the B-H and N-H 
surfaces that changes the position of surface states. In spite of that, the net charge transferred is not enough to open a band gap in 
the interface. Then, the N-H/B-H interface is metallic with free charge carries of opposite sign at each surface. The positive
charge will, therefore, attract the negative charge at adjacent surface, thereby producing a bidimensional electron gas. The
combination of bidimensional electron and hole gases, very close to each other, could lead to interesting transport properties. 

The applied uniaxial pressure decreases the distance between B-H and N-H surfaces at the interface that increases 
the overlap between orbitals of atoms at the surfaces. The left panel of Fig. 3 shows the band of gap of the interface as 
a function of the overlap volume between $s$
H orbitals\cite{overlap}. It is possible to see a precise linear dependence
of the band gap with the overlap, which corroborate the importance of overlap between orbitals of atoms at surface for the band gap of few-layered films. The value of pressure that corresponds to the greatest value of overlap in
Fig. 2 is about 70~Kbar, which can be achieved experimentally. Thus, any value of band gap ranging from zero up to 4.4~eV in the interface formed by the physical contact between B-H and N-H cBN surfaces can be tuned through the uniaxial applied pressure.
\begin{figure}
 \centering
\includegraphics[clip=true,scale=0.35]{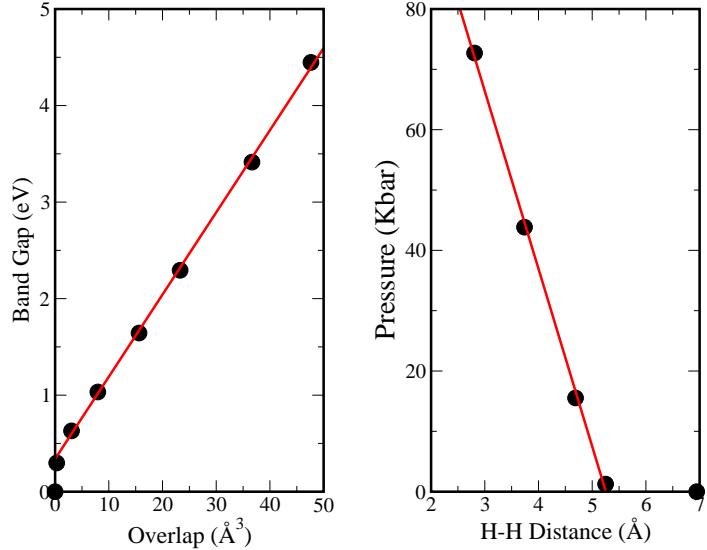}
 \caption{Left panel: band gap as a function of the overlap volume between hydrogen $s$ orbitals of H atoms at the interface formed
by the physical contact between B-H and N-H terminated cubic boron nitride, see left panel of Fig. 2.
Right panel: uniaxial pressure as a function the distance between hydrogen atoms at B-H terminated surface and N-H terminated surface.}
 \label{figure3}
\end{figure}

In summary, we report first-principles calculations on a tunable band gap interface composed of a high stable material, the cubic boron nitride.
Such result is very interesting in a applied point of view since: (i) high stable tunable semiconductors are highly desired for electronic
applications; (ii) the continuous dependence of the band gap with the applied pressure may be useful in very accurate electromechanical
devices; (iii) the spatial separation of the charge carries of opposite sign at the interface combined with the transparency of
cBN bulk for visible light may allow the production of photovoltaic devices. The mechanism whit leads to the dependence of the band gap
with applied pressure is understood in terms of the interaction between surface states. 

\begin{acknowledgments}
The authors acknowledge financial support from CNPq, FAPEMIG, Rede Nacional de Pesquisa em Nanotubos
de Carbono, and INCT-Nano-Carbono.
\end{acknowledgments}



\begin{thebibliography}{100}
\bibitem{overlap}
The overlap volume between two identical spheres is given by: $V=\frac{\pi}{12}(4R+d)(2R-d)^{2}$,
where $R$ is the sphere radius and $d$ is the distance between centers.
\end{thebibliography}
\end{document}